\documentstyle[aps,psfig,preprint]{revtex}
\begin{document}
~\\[-0.6in]
~\hspace*{3.9in} KSU-HEP-98-001\\
~\hspace*{3.9in} FNAL Pub-98/289-E\\
~\hspace*{3.9in} September 17, 1998 
\begin{center}
\Large
{\bf{
Measurement of the form-factor ratios for $D^+\rightarrow \overline{K}^{\,\star 0} 
\ell^+ \nu_\ell$
}}
\end{center}
%
%
\def\sameauthors#1{\hbox to\textwidth{\hss\vrule height.3cm width0pt\relax%
#1\hss}}

\large
\noindent
Fermilab E791 Collaboration\\[0.15in]
\normalsize
%
    E.~M.~Aitala,$^9$
       S.~Amato,$^1$
    J.~C.~Anjos,$^1$
    J.~A.~Appel,$^5$
       D.~Ashery,$^{14}$
       S.~Banerjee,$^5$
       I.~Bediaga,$^1$
       G.~Blaylock,$^8$
    S.~B.~Bracker,$^{15}$
    P.~R.~Burchat,$^{13}$
    R.~A.~Burnstein,$^6$
       T.~Carter,$^5$
 H.~S.~Carvalho,$^{1}$
  N.~K.~Copty,$^{12}$
    L.~M.~Cremaldi,$^9$
 C.~Darling,$^{18}$
       K.~Denisenko,$^5$
       A.~Fernandez,$^{11}$
       G.F.~Fox,$^{12}$
       P.~Gagnon,$^2$
       C.~Gobel,$^1$
       K.~Gounder,$^9$
     A.~M.~Halling,$^5$
       G.~Herrera,$^4$
 G.~Hurvits,$^{14}$
       C.~James,$^5$
    P.~A.~Kasper,$^6$
       S.~Kwan,$^5$
    D.~C.~Langs,$^{12}$
       J.~Leslie,$^2$
       B.~Lundberg,$^5$
       S.~MayTal-Beck,$^{14}$
       B.~Meadows,$^3$
       J.~R.~T.~de~Mello~Neto,$^1$
       D.~Mihalcea,$^7$
    R.~H.~Milburn,$^{16}$
 J.~M.~de~Miranda,$^1$
       A.~Napier,$^{16}$
       A.~Nguyen,$^7$
  A.~B.~d'Oliveira,$^{3,11}$
       K.~O'Shaughnessy,$^2$
    K.~C.~Peng,$^6$
    L.~P.~Perera,$^3$
    M.~V.~Purohit,$^{12}$
       B.~Quinn,$^9$
       S.~Radeztsky,$^{17}$
       A.~Rafatian,$^9$
    N.~W.~Reay,$^7$
    J.~J.~Reidy,$^9$
    A.~C.~dos Reis,$^1$
    H.~A.~Rubin,$^6$
    D.~A.~Sanders,$^9$
 A.~K.~S.~Santha,$^3$
 A.~F.~S.~Santoro,$^1$
       A.~J.~Schwartz,$^3$
       M.~Sheaff,$^{17}$
    R.~A.~Sidwell,$^7$
    A.~J.~Slaughter,$^{18}$
    M.~D.~Sokoloff,$^3$
    J.~Solano,$^1$
       N.~R.~Stanton,$^7$
    R.~J.~Stefanski,$^5$   
       K.~Stenson,$^{17}$
    D.~J.~Summers,$^9$
 S.~Takach,$^{18}$
       K.~Thorne,$^5$
    A.~K.~Tripathi,$^{7}$
       S.~Watanabe,$^{17}$
 R.~Weiss-Babai,$^{14}$
       J.~Wiener,$^{10}$
       N.~Witchey,$^7$
       E.~Wolin,$^{18}$
       D.~Yi,$^9$
    S.~M.~Yang,$^7$
       S.~Yoshida,$^{7}$                         
       R.~Zaliznyak,$^{13}$
       and
       C.~Zhang$^7$ \\
%
%
\small
\begin{center}
{\bf{Abstract}}
\end{center}
The form factor ratios $r_V = V(0)/A_1(0)$, $r_2 = A_2(0)/A_1(0)$ and $r_3 =
A_3(0)/A_1(0)$ in the decay $D^+ \rightarrow \overline{K}^{\,\star 0} \ell^+ \nu_\ell$,
$\overline{K}^{\,\star 0} \rightarrow K^- \pi^+$ have been measured
using data from charm hadroproduction experiment E791
at Fermilab. From 3034 (595) signal (background) events in the muon channel, we obtain
$r_V = 1.84 \pm 0.11 \pm 0.09$, $r_2 = 0.75 \pm 0.08 \pm 0.09$ 
and, as a first measurement of $r_3$, we find $0.04 \pm 0.33 \pm 0.29$. 
The values of the form factor ratios $r_V$ and $r_2$ measured for the
muon channel are combined with the values of $r_V$ and $r_2$ that we have
measured in the electron channel.
The combined E791 results for the muon and electron channels are
$r_V = 1.87 \pm 0.08 \pm 0.07$ and $r_2 =
0.73 \pm 0.06 \pm 0.08$.

\noindent
{\underline{~~~~~~~~~~~~~~~~~~~~~~~~~~~~~~~~~~}}\\

\small
\em
\noindent
$^1$ Centro Brasileiro de Pesquisas F\'\i sicas, Rio de Janeiro, Brazil\\
$^2$ University of California, Santa Cruz, California 95064\\
$^3$ University of Cincinnati, Cincinnati, Ohio 45221\\
$^4$ CINVESTAV, Mexico\\
$^5$ Fermilab, Batavia, Illinois 60510\\
$^6$ Illinois Institute of Technology, Chicago, Illinois 60616\\
$^7$ Kansas State University, Manhattan, Kansas 66506\\
$^8$ University of Massachusetts, Amherst, Massachusetts 01003\\
$^9$ University of Mississippi, University, Mississippi 38677\\
$^{10}$ Princeton University, Princeton, New Jersey 08544\\
$^{11}$ Universidad Autonoma de Puebla, Mexico\\
$^{12}$ University of South Carolina, Columbia, South Carolina 29208\\
$^{13}$ Stanford University, Stanford, California 94305\\
$^{14}$ Tel Aviv University, Tel Aviv, Israel\\
$^{15}$ Box 1290, Enderby, BC, V0E 1V0, Canada\\
$^{16}$ Tufts University, Medford, Massachusetts 02155\\
$^{17}$ University of Wisconsin, Madison, Wisconsin 53706\\
$^{18}$ Yale University, New Haven, Connecticut 06511\\

\noindent
{\underline{~~~~~~~~~~~~~~~~~~~~~~~~~~~~~~~~~~}}\\
\rm
\footnotesize
%
%
\narrowtext
\normalsize

The weak decays of hadrons containing heavy quarks are substantially influenced
by strong interaction effects.
Semileptonic charm decays such as
$D^+ \rightarrow \overline{K}^{\,\star 0} \ell^+ \nu_\ell$ are 
an especially clean way to study these effects because the 
leptonic and hadronic currents completely factorize in the decay amplitude.
All information about the
strong interactions can be parametrized by a few form factors.
Also, according to Heavy Quark
Effective Theory, the values of form factors for some semileptonic charm
decays can be related to those governing certain b-quark decays.
In particular, the form factors studied here can be related to those for the
rare $B$--meson decays $B \to K^\star e^+ e^-$
and $B \to K^\star \gamma$ \cite{key1,key2} which provide
windows for physics beyond the Standard Model.

With a vector meson in the 
final state, there are four form factors, $V(q^2)$, $A_1(q^2)$, $A_2(q^2)$
and $A_3(q^2)$, which are functions of the Lorentz-invariant
momentum transfer squared \cite{key3}. 
The differential decay rate for $D^+ \rightarrow \overline{K}^{\,\star 0} \mu^+
\nu_\mu$ with  $\overline{K}^{\,\star 0} \rightarrow K^- \pi^+$ is a quadratic
homogeneous function of the four form factors.
Unfortunately, the limited size of current data samples precludes precise
measurement of the $q^2$-dependence of the form factors; we thus assume the
dependence to be given by the
nearest-pole dominance
model: $F(q^2) = F(0)/(1-q^2/m_{pole}^2)$ where $m_{pole} = m_V = 2.1 \: {\rm
GeV}/c^2$ for the vector form factor $V$, and $m_{pole} = m_A = 2.5 \: {\rm GeV}/c^2$ for
the three axial-vector form factors \cite{key4}.
The third form factor $A_3(q^2)$, which is unobservable in the limit of
vanishing lepton mass, probes the spin-0 component of the off-shell $W$.
Additional spin-flip amplitudes, suppressed by an overall factor of
$m_{\ell}^2/q^2$ when compared with spin no-flip amplitudes,
contribute to the
differential decay rate.
Because $A_1(q^2)$ appears among the coefficients of every term in the
differential decay rate, it is customary to
factor out $A_1(0)$ and to measure the ratios $r_V =
V(0)/A_1(0)$, $r_2 = A_2(0)/A_1(0)$ and $r_3 = A_3(0)/A_1(0)$. The values of
these ratios can be extracted without any assumption about the total
decay rate or the weak mixing matrix element $V_{cs}$. 

We report new measurements of the form factor ratios for the muon channel
and combine them with slightly revised values of our previously published
measurements of $r_V$ and $r_2$ \cite{key5} for the electron channel. This
is the first set of measurements in both muon and electron channels from
a single experiment.
We also report the first
measurement of $r_3 = A_3(0)/A_1(0)$, which is unobservable in the limit of
vanishing charged lepton mass.

E791 is a fixed-target charm hadroproduction experiment \cite{key6}. Charm
particles
were produced in the collisions of a $500 \: {\rm GeV}/c$ $\pi^-$ beam with
five thin targets, one platinum and four diamond. About $2 \times 10^{10}$
events were recorded during the 1991-1992 Fermilab fixed-target run. The
tracking system consisted of 23 planes of silicon microstrip detectors, 45
planes of drift and proportional wire chambers, and two large-aperture dipole
magnets. Hadron identification is based on the information from two
multicell \v Cerenkov counters that provided good discrimination between kaons and
pions in the momentum range $6 - 36 \: {\rm GeV}/c$. In this momentum range, the
probabability of misidentifying a pion as a kaon depends on momentum but does
not exceed 5\%. We identified muon candidates using a single plane of
scintillator strips, oriented horizontally, located behind an equivalent of
2.4 meters of iron (comprising the calorimeters and one meter of bulk
steel shielding). The angular acceptance of the scintillator plane was $\approx\!\!\pm
62 \:{\rm mrad} \times \pm 48 \:{\rm mrad}$ (horizontally and vertically,
respectively), which is somewhat smaller than that of the rest of the spectrometer
for tracks which go through both
magnets ($\approx\!\pm 100 \:{\rm mrad} \times \pm 64 \:{\rm mrad}$).
The vertical position of a hit was
determined from the strip's vertical position, and the horizontal position 
of a hit from
timing information.

The event selection criteria used for this analysis 
are the same as for the electronic-mode form factor
analysis \cite{key5}, except for those related to lepton identification.
Events are selected if they contain an acceptable decay vertex
determined by the intersection point of three tracks
that have been identified as a muon, a kaon, and a pion. The longitudinal 
separation between this candidate decay vertex and the reconstructed production vertex
is required to be at least 15 times
the estimated error on the separation. The two hadrons must have opposite charge.
If the kaon and the muon have opposite charge, the event is assigned to the
``right-sign" sample; if they have the same charge, the event is assigned to the
``wrong-sign" sample used to model the background. 

To reduce the
contamination from hadron decays in flight, only muon candidates with
momenta larger than 8 GeV/$c$ are retained. 
With this momentum restriction, the efficiency of muon tagging was about 85\%, and
the probability for a hadron to
be identified as a muon was about 3\%.

To exclude feedthrough from $D^+ \rightarrow K^- \pi^+ \pi^+$,
we exclude events in which the invariant mass
of the three charged particles (with the muon candidate interpreted as a pion) is
consistent with the $D^+$ mass.
For our final selection criteria, we use a binary-decision-tree 
algorithm (CART \cite{key7}), which
finds linear combinations of parameters that have the highest discrimination
power between signal and background.
Using this algorithm, we found
a linear combination of four discrimination variables \cite{key5}:
(a) separation significance of the candidate decay vertex from target material;
(b) distance of closest approach of the candidate $D$ momentum vector to the
primary vertex, taking into account the maximum kinematically-allowed miss distance 
due to the unobserved neutrino; (c) product over candidate $D$ decay tracks of
the distance of closest approach of the track to the secondary vertex,
divided by the distance of closest approach to the primary vertex, where each distance 
is measured in units of measurement errors; and (d) significance of separation 
between the production and decay vertices.
This final selection criterion reduced the number of wrong-sign events by 50\%, and the number
of right-sign events by 25\%. 
Although this does not affect our sensitivity substantially, it does reduce 
systematic uncertainties associated with the background subtraction.

The minimum parent mass $M_{min}$ is defined as the invariant mass of $K \pi
\mu \nu$ when the neutrino momentum component along the $D^+$ direction of
flight is ignored.
The distribution of $M_{min}$ should have a Jacobian peak at the $D^+$
mass, and we observe such a peak in our data (Fig.~1).
We retain events with $M_{min}$ in the range 1.6 to 2.0 ${\rm
GeV}/c^2$ as indicated by the arrows in the figure.
The distribution of $K \pi$ invariant mass for the retained events
is shown in the top right of Fig.~1 
for both right-sign and wrong-sign samples. Candidates with $0.85 < M_{K \pi}
< 0.94 \: {\rm GeV}/c^2$ were retained, yielding final data samples of 3629
right-sign and 595 wrong-sign events.

The hadroproduction of charm, the differential decay rate, and the detector response were simulated
with a Monte Carlo event generator. A sample of events was generated
according to the differential decay rate (Eq.~22 in Ref.~\cite{key3}), with the
form factor ratios
$r_V = 2.00$, $r_2 = 0.82$,
and $r_3 = 0.00$.
The same selection criteria were applied to the Monte Carlo events 
as to real data.
Out of 25 million generated events, 95579 decays passed all cuts.
Figure~1 (bottom) shows the distribution of $M_{K \pi}$ from real data after background
subtraction (``right-sign" minus ``wrong-sign") 
overlaid with the corresponding Monte Carlo distribution after
all cuts are applied. The agreement between the two distributions suggests 
that wrong-sign events correctly account for the size of the background.

The differential decay
rate \cite{key3} is expressed in terms of four independent kinematic variables:
the square of the momentum transfer ($q^2$), the polar
angle $\theta_V$ in the $\overline{K}^{\,\star 0}$ rest frame
between the $\pi^+$ and $D^+$, the polar angle $\theta_\ell$ in the $W^+$ rest frame
between the $\nu_\mu$ and $D^+$,
and the azimuthal angle $\chi$ in the $D^+$
rest frame between the $\overline{K}^{\,\star 0}$ and $W^+$ decay planes.
The definition we use for the polar angle $\theta_\ell$ is related to
the definition used in Ref. \cite{key3} by $\theta_\ell \to \pi - \theta_\ell$.

Semileptonic decays cannot be fully reconstructed due to the undetected
neutrino. With the available information about the $D^+$ direction of flight and
the charged daughter particle momenta, the neutrino momentum 
(and all the decay's kinematic variables) can be determined up
to a two-fold ambiguity if the parent mass is constrained. Monte Carlo
studies show that the differential decay rate is more accurately determined
if it is calculated 
with the solution corresponding to the lower
laboratory-frame neutrino momentum.

To extract the form factor ratios the distribution of the data points in
the four-dimensional kinematic variable space is fit to the full
expression for the differential decay rate.
We use the same unbinned maximum-likelihood fitting technique as
in our $D^+ \rightarrow \overline{K}^{\,\star 0} e^+ \nu_e$ form factor analysis
\cite{key5}.
The likelihood function is computed from the density of weighted Monte
Carlo events 
(described above) near each data event 
in the four-dimensional space of kinematic
variables \cite{key8}. 
To include background in the fit, a similar likelihood function based on
the density of wrong-sign events around each right-sign event is used.
With this method the fitted results are
subject to small systematic biases which originate from two sources: (a)
approximate normalization of the likelihood function;
(b) nonlinearity of the decay rate within the volume centered on the data point.
These systematic biases of the fitted parameters were determined
from Monte Carlo studies, and are $\delta r_V = +0.09 \pm 0.02$, $\delta r_2 = -0.08
\pm 0.01$ and $\delta r_3 = -0.11 \pm 0.06$. After correction for these biases
by subtracting these $\delta r$\hspace{-0.01cm}'s from the measured values, the final
form factor ratios and their statistical errors are $r_V = 1.84 \pm
0.11$, $r_2 = 0.75 \pm 0.08$ and $r_3 = 0.04 \pm 0.33$.
The sensitivity of the fit to $r_3$ is low
because this form factor ratio only contributes to
spin-flip amplitudes. The correlation coefficient between the form factors
to which we are most sensitive,
$r_V$ and $r_2$, is $-0.090$. The correlation
coefficient between $r_3$ and $r_2$ is $-0.211$, and between $r_3$
and $r_V$ is $-0.087$.
Figure~2 compares the data and Monte Carlo distributions in various regions
of the four-dimensional phase space.

We checked for any potential bias in these results due to our choice of neutrino
momentum by employing a secondary fitting technique.
Again, Monte Carlo is used to account for detector acceptance and smearing.
However, both solutions for the neutrino momentum are now used in the fit.
We divide the four-dimensional kinematic variable space into 240 separate
volumes and determine the number of data entries in each volume, where
each event has two entries -- one for each neutrino-momentum solution. We
use Monte Carlo events to determine the probabilities that an event generated
in a particular phase space volume will be observed in each of the other
volumes when the wrong neutrino-momentum choice is used. This feedthrough
probability matrix and the observed number of data events determine the
fraction of data events that correspond to the correct neutrino-momentum
solution in each volume of kinematic variable space. Each fraction is
then used in a binned maximum likelihood fit. Background is modeled with
wrong-sign candidates as in the primary method.
The form factor ratios and statistical errors measured
with this secondary method are $r_V = 1.90 \pm 0.11$, $r_2 =
0.72 \pm 0.08$, and $r_3 = -0.25 \pm 0.34$.
This method for extracting the form factor ratios uses the same
$D^+ \to \overline{K}^{\,\star 0} \mu^+ \nu_\mu$ candidates as the previous method,
but uses additional neutrino-momentum solutions. This is true for both the 
data and for the Monte Carlo sample used in the likelihood function calculation,
so the results of this fit could differ from those of the
previous fit. 

The values of the form factor ratios obtained with the two
methods agree well, providing further assurance that selecting the lower
neutrino momentum solution in the primary method and correcting for the
systematic bias gives the correct result. However, the systematic uncertainties
for the primary method (see below) were found to be significantly smaller,
mainly because the unbinned maximum-likelihood method is more stable against
changes in the size of the phase space volume. Therefore, the 
primary method was chosen for quoting final results.

We classify systematic uncertainties into three categories: (a) Monte Carlo
simulation of detector effects and production mechanism; (b) fitting
technique; (c) background subtraction. The estimated contributions of each are given
in Table~\ref{tabone}. The main contributions to category (a) are
due to muon identification and data selection criteria.
The contributions to category (b) are related to the limited size
of the Monte Carlo sample
and to corrections for systematic bias. 

The measurements of the form factor ratios for $D^+ \rightarrow \overline{K}^{\,\star
0} \mu^+ \nu_\mu$ presented here and for the similar decay channel $D^+
\rightarrow \overline{K}^{\,\star 0} e^+ \nu_e$ \cite{key5} follow the same analysis
procedure except for the charged lepton identification. Both results
are listed in Table~\ref{tabtwo}.
The consistency within errors of the results measured in the electron
and muon channels 
supports the assumption that strong interaction effects, incorporated
in the values of form factor ratios, do not depend on the particular
$W^+$ leptonic decay. Based on this assumption, we combine the results
measured for the electronic and muonic decay modes.
The averaged values of the form factor ratios are $r_V =
1.87 \pm 0.08 \pm 0.07$ and $r_2 = 0.73 \pm 0.06 \pm 0.08$. 
The statistical and systematic uncertainties of the average results
were determined using the general procedure described in Ref. \cite{key9}
(Eqns.~3.40 and~3.40${}^{\prime}$). Some of the systematic errors 
for the two samples have positive correlation coefficients, and some
negative.
The combination of all systematic errors is ultimately close to that which one
would obtain assuming all the errors are uncorrelated.
The third form
factor ratio $r_3$ was not measured in the electronic mode.

Table~\ref{tabtwo} compares the values of the form factor ratios $r_V$ and $r_2$
measured by E791 in the electron, muon and combined modes with previous
experimental results.
The size of the data sample and the decay channel are
listed for each case. All experimental results are consistent 
within errors. The comparison between the E791 combined values of the form 
factor ratios $r_V$ and $r_2$ and previous experimental results
is also shown in Fig.~3 (top).
Table~\ref{tabthree} and Fig.~3 (bottom) compare the final E791 result with 
published theoretical predictions.
The spread in the theoretical
results is significantly larger than the E791 experimental errors.

To summarize, we have measured the values of the form factor ratios 
in the
decay channel $D^+ \rightarrow \overline{K}^{\,\star 0} \mu^+ \nu_\mu$ to be
$r_V = 1.84 \pm 0.11 \pm 0.09$, $r_2 = 0.75 \pm 0.08 \pm 0.09$ and
$r_3 = 0.04 \pm 0.33 \pm 0.29$. The data sample has about 3000 events
after subtracting 595 background events. Combining these results for
$r_V$ and $r_2$ with those measured by E791 for the decay channel $D^+
\rightarrow \overline{K}^{\,\star 0} e^+ \nu_e$ gives
$r_V = 1.87 \pm 0.08 \pm 0.07$ and $r_2 = 0.73 \pm 0.06 \pm 0.08$.

We gratefully acknowledge the assistance from Fermilab and
other participating institutions. This work was supported by the
Brazilian Conselho Nacional de Desenvolvimento Cient\'\i fico e
Technol\'{o}gico, CONACyT (Mexico), the Israeli Academy of Sciences and
Humanities, the U.S. Departament of Energy, the U.S.-Israel Binational
Science Foundation, and the U.S. National Science Foundation.

\newpage
\begin{table}[tbh]
\caption{The main contributions to uncertainties on the form factor ratios.}
\label{tabone}
\begin{center}
\begin{tabular}{lccc}
{\bf Source} & $\sigma_{r_2}$ & 
$\sigma_{r_V}$ & $\sigma_{r_3}$ \\ \hline
{\bf Simulation of detector effects:} & 0.06 & 0.08 & 0.16  \\ \hline
$\;\;\;$Hadron identification & 0.01 & 0.01 & 0.02 \\
$\;\;\;$Muon identification & 0.04 & 0.06 & 0.10 \\
$\;\;\;$Production mechanism & 0.01 & 0.01 & 0.02 \\
$\;\;\;$Acceptance & 0.03 & 0.02 & 0.08 \\
$\;\;\;$Cut selection & 0.03 & 0.04 & 0.09 \\ \hline
{\bf Fitting technique:} & 0.02 & 0.03 & 0.22 \\ \hline
$\;\;\;$MC volume size & 0.02 & 0.02 & 0.12 \\
$\;\;\;$Number of MC points & 0.01 & 0.01 & 0.18 \\
$\;\;\;$Bias & 0.01 & 0.02 & 0.06 \\ \hline
{\bf Background:} & 0.06 & 0.04 & 0.08 \\ \hline
$\;\;\;$No. of background events & 0.04 & 0.02 & 0.06 \\
$\;\;\;$Background shape & 0.04 & 0.04 & 0.06 \\ \hline
{\bf Total} & 0.09 & 0.09 & 0.29 \\
\end{tabular}
\end{center}
\end{table}

\begin{small}
\begin{table}[tbh]
\caption{Comparison of E791 results with previous experimental results.
The E791 electron result for $r_V$ is 0.06 higher than
the value reported in Ref.~[5] because we have corrected for
inaccuracies in the earlier modeling of the $D^+$ transverse momentum.}
\label{tabtwo}
\begin{center}
\begin{tabular}{llll}
Exp. & Events & $r_V = V(0)/A_1(0)$ & $r_2 = A_2(0)/A_1(0)$  \\ \hline 
E791 & 6000 ($e+\mu$) & $1.87 \pm 0.08 \pm 0.07$ & $0.73 \pm 0.06 \pm 0.08$ \\
E791 &  3000 ($\mu$) & $1.84 \pm 0.11 \pm 0.09$ & $0.75 \pm 0.08 \pm 0.09$ \\
E791 & 3000 ($e$) & $1.90 \pm 0.11 \pm 0.09$ & $0.71 \pm 0.08 \pm 0.09$ \\
E687\cite{key10} & 900 ($\mu$) & $1.74 \pm 0.27 \pm 0.28$ & $0.78 \pm 0.18 \pm 0.10$ \\
E653\cite{key11} & 300 ($\mu$) & $2.00^{+0.34}_{-0.32} \pm 0.16$ & $0.82^{+0.22}_{-0.23} \pm 0.11$ \\
E691\cite{key12} & 200 ($e$) & $2.0 \pm 0.6 \pm 0.3$ & $0.0 \pm 0.5 \pm 0.2$ \\ 
\end{tabular}
\end{center}
\end{table}
\end{small}

\begin{table}[tbh]
\caption{Comparison of E791 results with theoretical predictions for the form
factor ratios $r_V$ and $r_2$.}
\label{tabthree}
\begin{center}
\begin{tabular}{lll}
Group                &        $r_V$                  &       $r_2$       \\ \hline
E791 ($e$ and $\mu$) &    $1.87 \pm 0.11$      & $0.73 \pm 0.10$ \\ \hline
ISGW2\cite{key13}      &        $2.0$                  &           $1.3$  \\ 
WSB\cite{key14}       &        $1.4$                  &           $1.3$  \\
KS\cite{key3}       &        $1.0$                  &           $1.0$ \\
AW/GS \cite{key15,key16} &        $2.0$                  &           $0.8$ \\ 
Stech\cite{key17}    &        $1.55$                 &           $1.06$ \\
BKS\cite{key18}      &     $1.99 \pm 0.22 \pm 0.33$  &    $0.70 \pm 0.16 \pm 0.17$ \\    
LMMS\cite{key19}     &        $1.6 \pm 0.2$          &     $0.4 \pm 0.4$ \\
ELC \cite{key20}     &        $1.3 \pm 0.2$          &     $0.6 \pm 0.3$ \\
APE \cite{key21}     &        $1.6 \pm 0.3$          &     $0.7 \pm 0.4$ \\
UKQCD \cite{key22} &     $1.4_{-0.2}^{+0.5}$ &    $0.9 \pm 0.2$ \\
BBD \cite{key23}    &        $2.2 \pm 0.2$          &     $1.2 \pm 0.2$ \\
LANL \cite{key24}   &   $1.78 \pm 0.07$      &  $0.68 \pm 0.11$ \\
\end{tabular}
\end{center}
\end{table}

\begin{figure}[tbh]
\centerline{
\psfig{figure=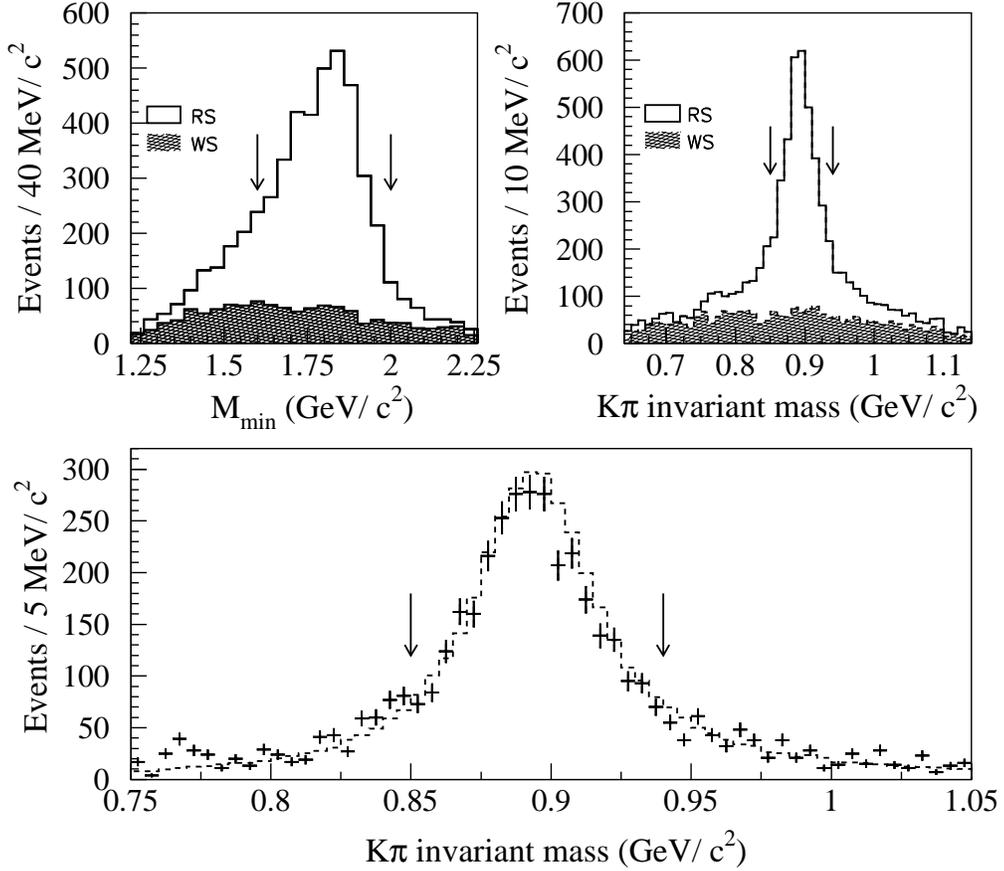,width=5.5in}}

\caption{Distributions of minimum parent mass $M_{min}$ and $K \pi$ invariant
mass for $D^+ \rightarrow \overline{K}^{\,\star 0} \mu^+ \nu_\mu$ candidate events.
Right-sign (RS) and wrong-sign (WS) samples are defined in the text.
Top left: $M_{min}$ for events with $K \pi$ mass in the range
0.85 to 0.94 GeV/$c^2$. Top right: $K \pi$ invariant mass
for events with $M_{min}$ in the range 1.6 to 2.0 GeV/$c^2$.
Bottom: background-subtracted (RS-WS)  $K \pi$ mass distribution (crosses)
compared to Monte Carlo prediction (dashed histogram) for
events with $M_{min}$ in the range 1.6 to 2.0 GeV/$c^2$.
All candidates pass all the 
other final selection cuts. The arrows indicate the range of the final sample.}
\end{figure}

\begin{figure}[tbh]
\centerline{
\psfig{figure=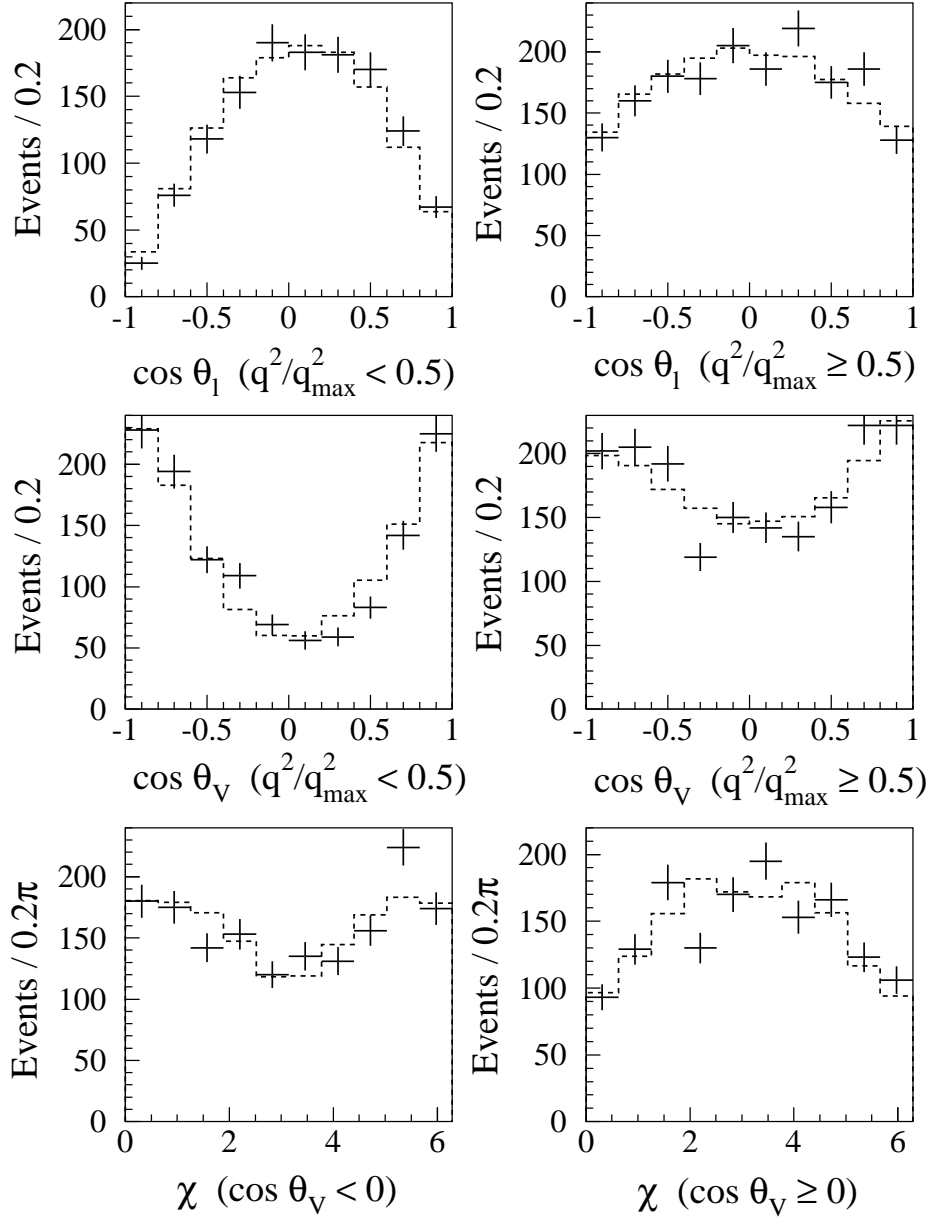,width=5.0in}}

\caption{Comparison of single-variable distributions of background-subtracted
data (crosses) with Monte Carlo predictions (dashed histograms) using
best-fit values for the form factor ratios.}
\end{figure}

\begin{figure}[tbh]
\vspace*{-1.3in}
\centerline{
\psfig{figure=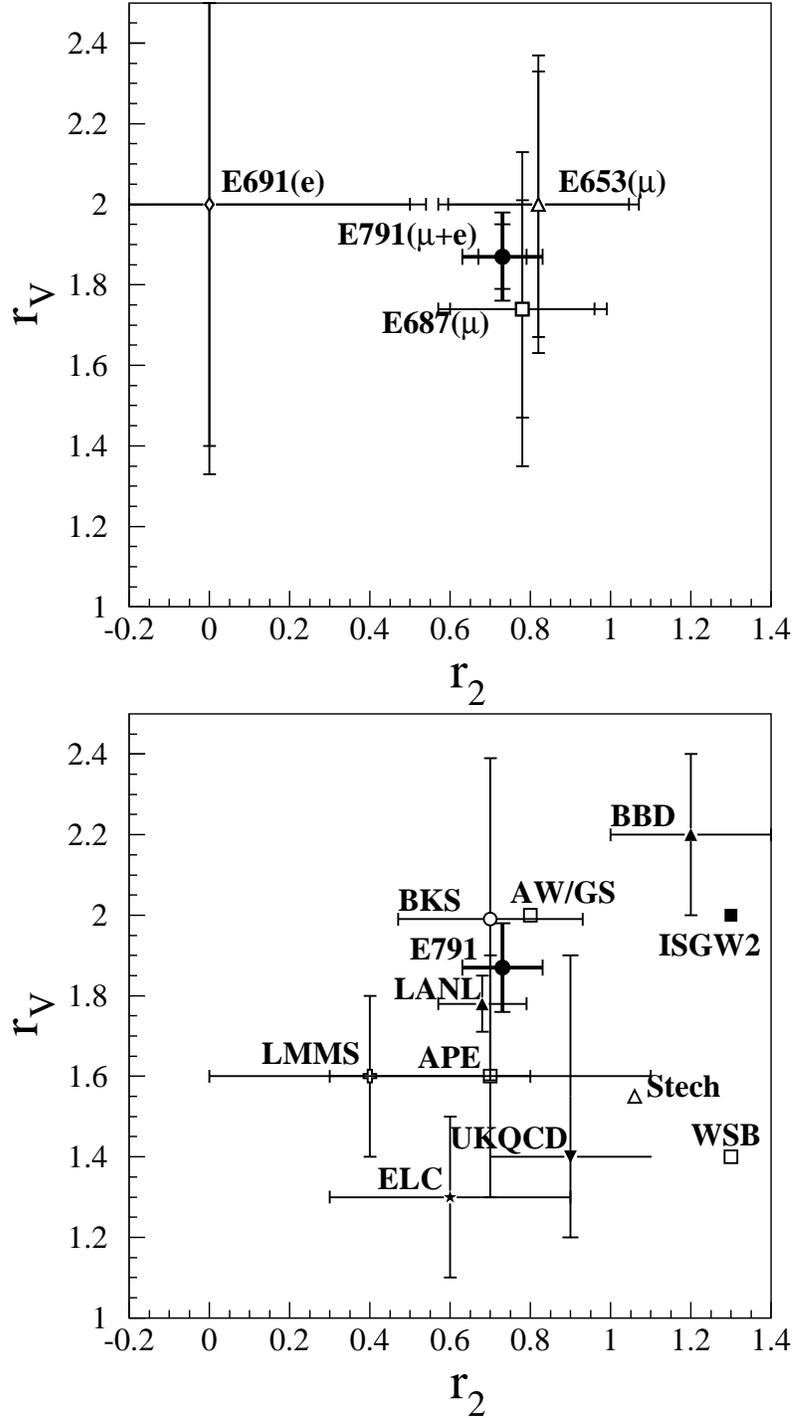,width=4.2in}}

\caption{Top: Comparison of experimental measurements of form factor
ratios $r_V$ and $r_2$ for $D^+ \rightarrow \overline{K}^{\,\star 0} \ell^+ \nu_\ell$
in the muon ($\mu$), electron ($e$) and combined ($\mu + e$) channels.
The smaller error bars indicate the statistical errors and the larger
ones indicate the statistical and sytematic errors added in quadrature.
Bottom: Comparison of theoretical predictions with the E791 ($\mu+e$)
result.}
\end{figure}

\end{document}